\documentclass[epj]{svjour}
\usepackage{epsf}
\usepackage{latexsym}
\begin{document}
\date{}
\title{Effect of a lattice upon an interacting system of electrons 
in two dimensions: Breakdown of scaling and decay of persistent currents}

\titlerunning{Effect of a lattice}

\author{H. Falakshahi $^{(a)}$, Z. \'A. N\'emeth$^{(a,b)}$ and J.-L. 
Pichard$^{(a,c)}$}

\authorrunning{H. Falakshahi et al}
        
\institute{
(a) CEA/DSM, Service de Physique de l'Etat Condens\'e, 
Centre d'Etudes de Saclay, 91191 Gif-sur-Yvette Cedex, France \\  
(b) E\"otv\"os University, Departement of Physics of Complex Systems, 
1117 Budapest, P\'azm\'any P\'eter s\'et\'any 1/A, Hungary \\
(c) Laboratoire de Physique Th\'eorique et Mod\'elisation, 
Universtit\'e de Cergy-Pontoise, 95031, Cergy-Pontoise Cedex, France 
}

\abstract{The ground state of an electron gas is characterized 
by the interparticle spacing to the effective Bohr radius ratio 
$r_s=a/a_B^{*}$. For polarized electrons on a two dimensional 
square lattice with Coulomb repulsion, we study the threshold 
value $r_s^*$ below which the lattice spacing $s$ becomes a 
relevant scale and $r_s$ ceases to be the scaling parameter. 
For systems of small ratios $s/a_B^{*}$, $s$ becomes only relevant at 
small $r_s$ (large densities) where one has a quantum fluid 
with a deformed Fermi surface. For systems of large $s/a_B^{*}$, 
$s$ plays also a role at large $r_s$ (small densities) 
where one has a Wigner solid, the lattice limiting its harmonic 
vibrations. The thermodynamic limit of physical systems of different 
$a_B^{*}$ is qualitatively discussed, before quantitatively 
studying the lattice effects occurring at large $r_s$. Using a few 
particle system, we compare exact numerical results 
obtained with a lattice and analytical perturbative expansions 
obtained in the continuum limit. Three criteria giving similar 
values for the lattice threshold $r_s^*$ are proposed. The first 
one is a delocalization criterion in the Fock basis of lattice 
site orbitals. The second one uses the persistent current which 
can depend on the interaction in a lattice, while it becomes independent 
of the interaction in the continuum limit. The third one 
takes into account the limit imposed by the lattice to the harmonic 
vibrations of the electron solid.  
} 
 
 
\maketitle

\section{Introduction}

 When one considers interacting electrons free to move in an immobile 
background of positive ions, one can represent the ions by a uniform 
positive jellium if the electron density is sufficiently small. This  
uniform jellium gives simply rise to a constant term in the Hamiltonian. 
One gets a continuum model characterized by two  scales: the inter-electron 
spacing $a$ and the effective Bohr radius $a_B^{*}$. Simple scaling laws 
are obtained if one uses the dimensionless ratio $r_s=a/a_B^{*}$. 
This continuum approximation neglects the discrete character of the 
lattice of positive ions. 

If one wants to keep the periodic character of the ionic lattice, 
one has to include a periodic potential instead of a uniform jellium 
or to use the tight-binding approximation. One obtains a lattice 
model, where the kinetic energy can be simplified if the hopping 
terms are restricted to nearest neighbor ions. A lattice introduces 
a third scale: the lattice spacing $s$. If $s$ is irrelevant, the 
lattice model keeps the same universal scaling than the continuum 
limit, if one uses the combination of lattice parameters which 
becomes $r_s=a/a_B^{*}$ in the continuum.

 We study when the low energy spectrum of the lattice model can be 
described by a continuum approximation, the lattice effects remaining 
only important in the high energy spectrum. What is the carrier density 
above which the scale $s$ becomes relevant and $r_s$ ceases to be the 
scaling parameter for the lattice ground state? The answer depends on $s$ 
and on the two parameters controlling the effective Bohr radius   
\begin{equation}
a_B^{*}=\frac{\varepsilon_r \hbar^2}{m^* e^2}:
\end{equation}
the effective mass $m^*$ of the carriers and the dielectric constant 
$\varepsilon_r$ of the medium. 

If $a_B^{*}$ is large compared to $s$, the lattice effects are only 
important for small values of $r_s$, for which one has a quantum fluid 
with a deformed Fermi surface. This is a highly quantum weak coupling 
limit of large carrier densities. If the issue is to study charge 
crystallization in such systems, the densities of interest are much 
lower than those required to deform the Fermi surface, and the physics 
can be described in the continuum approximation. If one uses a 
$L \times L$ lattice model with $N$ particles to study electron  
crystallization for systems of small $s/a_B^{*}$, one has to take 
lattice fillings $N/L^2$ and hence ratios $r_s$ above the lattice 
threshold $r_s^*$, where the lattice model can be described by the 
continuum limit. 

If $a_B^{*}$ is small compared to $s$, the lattice effects become 
important when $r_s$ is large also. This is a limit where the lattice 
physics takes place also in a weakly quantum strong coupling regime, 
at low densities. If one continues to increase the density in those 
systems, one can eventually reach the limit usually described by a  
Hubbard model near half filling, where the lattice can give rise 
to a Mott insulator if the interaction is large enough. In contrast 
to the case where $a_B^{*}$ is large compared to $s$, the continuum 
approximation cannot be assumed for studying electron crystallization. 
One has the obvious problem of commensurability between the electron 
lattice characterizing the continuum limit and the ionic lattice. Even 
if there is commensurability, there is a remaining limit imposed by the 
lattice to the harmonic vibrations of the Wigner solid. 

Eventually, let us note that for a lattice model, it is important to 
know the electron density below which its low energy spectrum begins 
to exhibit the continuum behavior and its universal scaling laws, if 
one uses the combination of lattice parameters which becomes 
$r_s=a/a_B^{*}$ in the continuum.
 
\section{Two dimensional continuum model}
\label{section1}

 The Hamiltonian $H_c$ describing  $N$ polarized electrons of mass 
$m$ free to move on a continuum space of dimension $d$ and dielectric 
constant $\varepsilon_r=1$ contains one body kinetic terms, two body 
interaction terms plus the constant term due to the presence of the 
uniform background of positive ions necessary to have charge neutrality. 
\begin{equation}
H_c=-\frac{\hbar^2}{2m} \sum_{i=1}^N \nabla_i^2 + e^2 
\sum_{1\leq i < j\leq N} \frac{1}{|{\bf r}_i-{\bf r}_j|} + const, 
\end{equation}
Measuring the energies in rydbergs ($1 Ry = me^4/2\hbar^2$) and 
the lengths in units of the radius $a$ of a sphere (circle in $2d$) 
which encloses on the average one electron, $e$ and $m$ being the 
electronic charge and mass, $H_c$ becomes 
\begin{equation}
H_c=-\frac{1}{r_s^2} \sum_{i=1}^N \nabla_i^2 + \frac{2}{r_s}  
\sum_{1\leq i < j\leq N} \frac{1}{|{\bf r}_i-{\bf r}_j|} + const, 
\label{H-continuous}
\end{equation}
where $r_s=a/a_B$. The Bohr radius $a_B=\hbar^2/me^2$ is a measure of 
the GS radius of the hydrogen atom while the rydberg ($ 1Ry=e^2/(2a_B)$) 
is its GS binding energy. Eq. \ref{H-continuous} tells us that the 
physics of a system of interacting electrons in the continuum does not 
depend on many independent parameters ($\hbar$, $e$, $m$, the electronic 
density $n_s$...) but only on a single dimensionless scaling ratio 
$r_s=a/a_B$ when $N \rightarrow \infty$, $a_B$ characterizing the scale 
for the quantum effects. 

For the GS, if many electrons are inside the quantum volume $a_B^d$, 
one gets the weak coupling limit (small $r_s$) where one has a Fermi 
liquid.  Though our theory could be easily extended to arbitrary 
dimensions and could include the spin degrees of freedom, we restrict 
the study  to polarized electrons (spinless fermions) and to the 
dimension $d=2$. The ground state (GS) energy in rydbergs 
per particle is given in the Hartree-Fock approximation \cite{brueckener} 
as: 
\begin{equation} 
E_0=\frac{h_0}{r_s^2}+\frac{h_1}{r_s}+O(\ln r_s), \ \ r_s \ll 1
\label{weak-coupling}
\end{equation}
with coefficients $h_0=2$ for the kinetic energy and $h_1=-1.6972$ for  
the exchange energy. 

In the strong coupling limit (large $r_s$), the volume per electron 
$a^d$ is large compared to $a_B^d$, and the electrons crystallize on 
an hexagonal lattice with weak quantum effects (Wigner crystal). As 
Wigner's original approximation\cite{wigner} suggests, the GS energy 
per particle in rydbergs can be expanded in powers of $r_s^{1/2}$ :  
\begin{equation} 
E_0=\frac{f_0}{r_s}+\frac{f_1}{r_s^{3/2}}+\frac{f_2}{r_s^{2}}+O(r_s^{5/2})  
, \ \ r_s \gg 1
\label{strong-coupling}
\end{equation}
The leading term ($\propto r_s^{-1}$) is of classical nature (Coulomb 
energy of the lattice of electrons in a continuum background of positive 
charge) while the first correction ($\propto r_s^{-3/2}$) is quantum 
(zero point energy of the harmonic oscillations of the electrons about 
their lattice points). One gets \cite{carr,bonsall,ceperley1} 
$f_0=-2.2122$ and $f_1=1.628$ in two dimensions.    

 Using quantum Monte Carlo methods, the two dimensional crossover 
between these two limits has been studied. A variational approach 
\cite{ceperley1} and a Green function Monte Carlo approach 
\cite{ceperley2} have given a critical ratio $r_s^c \approx 37$ 
for a possible transition separating the quantum fluid from the Wigner 
solid in the continuum. However, the well-known sign problem of the 
Monte Carlo methods requires to impose the nodal structures 
of the solutions, making this picture not free of certain 
assumptions.

  Assuming periodic boundary conditions (BCs) for $N$ polarized 
electrons in a square of size $D$, one can ignore the constant term 
in $H_c$, the electronic density $n_s=N/D^2$ and $a=1/\sqrt{\pi n_s}$.

\section{Square lattice model}
\label{section2}

 We now define a square lattice model of spacing $s$, size $L=D/s$, 
nearest neighbor hopping element 
\begin{equation}
t=\frac{\hbar^2}{2m s^2}
\end{equation}
and interaction strength 
\begin{equation}
U=\frac{e^2}{s}. 
\end{equation}
The lattice Hamiltonian $H_l$ reads:   
\begin{equation}
H_l=t \left(4N- \sum_{\left<{\bf j},{\bf j'}\right>}
c_{\bf j}^{\dagger} c_{\bf j'}\right) + \frac{U}{2}  
\sum_{{\bf j} \neq {\bf j'}} \frac{n_{\bf j} n_{\bf j'}}
{|d_{\bf jj'}|}.  
\label{H-lattice-site}
\end{equation}
The operators $c_{\bf{j}}^{\dagger}$ ($c_{\bf{j}}$) create 
(annihilate) a polarized electron (spinless fermion) at the site 
$\bf{j}$ and $\left< {\bf j},{\bf j'}\right>$ means that the sum 
is restricted to nearest neighbors. $d_{\bf jj'}$ is the distance  
between the sites ${\bf j}$ and ${\bf j'}$ in unit of $s$. 

The Hamiltonian (\ref{H-lattice-site}) can also be written using the 
operators $d_{\vec{k}}^{\dagger}$ ($d_{\vec{k}}$) creating (annihilating) 
a polarized electron in a plane wave state of momentum $\vec{k}$: 
\begin{eqnarray}
H_l &=&4Nt - 2t \sum_k \left(\cos k_x +\cos k_y \right)d_{\bf k}^\dagger
d_{\bf k} \nonumber \\ 
&&+U \sum_{{\bf k},{\bf k'},{\bf q}} V({\bf q}) d_{{\bf k}+{\bf q}}^\dagger 
d_{{\bf k'}-{\bf q}}^\dagger d_{\bf k'} d_{\bf k}
\label{Hamiltonian-lattice-k}
\end{eqnarray}
where
\begin{equation}
V({\bf q})={1\over 2 L^2} \sum_{\bf j} {\cos{\bf qj}\over d_{\bf j0}}. 
\label{int}
\end{equation}
The states of different total momenta $\vec{K}$ are decoupled. 
In the lattice units, $1 Ry=U^2/4t$, $a_B=2st/U$ and the 
ratio $r_s$ becomes:      
\begin{equation}
r_s=\frac{a}{a_B}=\frac{UL}{2t\sqrt{\pi N}}.
\label{r_s-lattice}
\end{equation}  

\begin{figure}
\begin{center}
\epsfxsize=7cm 
\epsfbox{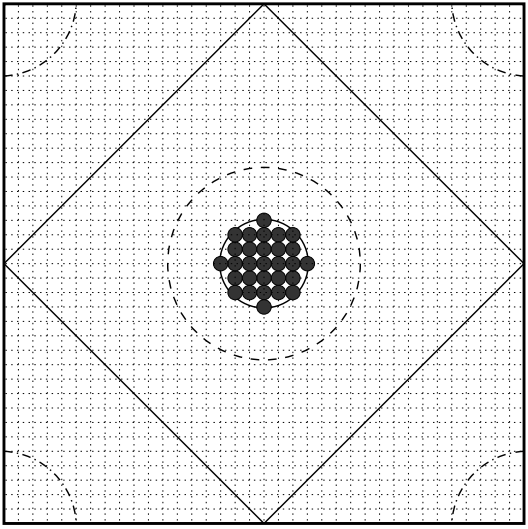}
\end{center}
\caption{Weak coupling limit: GS occupation numbers of a 
non interacting system in the reciprocal lattice. 
The Fermi-surfaces are sketched for increasing numbers of 
particles in a $36 \times 36$ square lattice. At low 
fillings, the Fermi surface is almost a circle, while 
it becomes deformed at larger fillings.}
\label{Fig1}
\end{figure}

\begin{figure}
\begin{center}
\epsfxsize=7cm 
\epsfbox{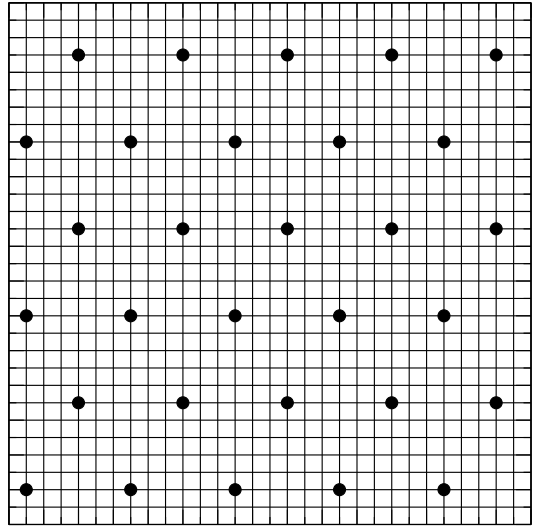}
\end{center}
\caption{Strong coupling limit: GS occupation numbers 
of a Wigner solid in real space. The hexagonal Wigner lattice 
of $N$ electrons becomes commensurate with the 
$L \times L$ square lattice if one takes $N=L=30$.}
\label{Fig2}
\end{figure}

\section{Lattice effects in the thermodynamic limit}
\label{section3}

\subsection{Lattice threshold in the weak coupling limit}

 In the limit $r_s \rightarrow 0$, the GS energy is 
mainly kinetic. This is a consequence of the Pauli exclusion 
principle, which pushes the Fermi energy of the non interacting 
system to much higher values than the classical Coulomb energy.  
The GS kinetic energies of $H_l$ and $H_c$ become different when 
the Fermi surface is deformed by the lattice. Such a deformation 
can give rise to nesting effects with important consequences 
discussed in the literature \cite{gonzales,guinea,schulz}.  
The Fermi wave vectors ${\bf k_F}$ are  given by 
$4t - 2t (\cos k^F_x + \cos k^F_y) = \epsilon_F$ for a square 
lattice of Fermi level $\epsilon_F$, instead of 
$t (k^F_x+k^F_y)^2= \epsilon_F$ for the continuum limit. 

Expanding the kinetic part of the lattice Hamiltonian 
(\ref{Hamiltonian-lattice-k}) for small wave vectors, one gets:
\begin{equation}
H_l^{kin} \approx t \sum_{k<k_F} {\bf k}^2 - {t \over 12} 
\sum_{k<k_F} \left( k_x^4 + k_y^4 \right)
\end{equation}
For the first term, one recovers the kinetic term of the continuum 
expansion (\ref{weak-coupling}): 
\begin{eqnarray}
 t \sum_{k < k_F} {\bf k}^2 &\approx& \int {D^2 \over (2\pi)^2} 
{\hbar^2 \over 2m} {\bf k}^2 d^2{\bf k} \nonumber \\
&=& {N^2 \hbar^2 \pi \over m D^2} = {2 \over r_s^2} (N Ry), 
\label{E_K}
\end{eqnarray}
while the lattice correction reads:
\begin{eqnarray}
\Delta E_{l} &=& - {t \over 12} \sum_{\bf k} k_x^4 + k_y^4 \nonumber 
\\ &=& - {\hbar^2 \over 24 m s^2} \int {L^2 \over (2\pi)^2} 
(k_x^4 + k_y^4)d^2{\bf k} 
\end{eqnarray}
which becomes, using $k_F = \sqrt {4N \pi}/ L$:
\begin{equation}
\Delta E_{l} = - {\hbar^2 N^3 \pi^2 \over m L^4 s^2} = -N Ry {2\over
r_s^2} {\pi N\over L^2}.
\label{DeltaE}
\end{equation}
The condition for having the lattice correction $\Delta E_{l}$ 
(Eq. \ref{DeltaE}) smaller than the continuum kinetic energy 
(Eq. \ref{E_K}) yields 
\begin{equation}
r_s > r_s^* \approx {s \over a_B}. 
\end{equation}
This estimate of $r_s^*$ is only valid when $r_s \rightarrow 0$, 
since it neglects  the effect of the interaction. When one turns 
on the interaction, transitions from states below the Fermi surface 
to states above it (see Hamiltonian (\ref{Hamiltonian-lattice-k}) 
and Fig. \ref{Fig1}) take place. This smears the Fermi surface, 
giving an uncertainty $\Delta k_F$ to  $k_F$. This uncertainty is 
evaluated in Appendix \ref{Appendix A}. One gets:   
\begin{equation}
\Delta k_F = {U^2 L\over t^2  \sqrt{N}} {I^2\over 16 \sqrt{4 \pi}}. 
\end{equation}
Since $\cos(k_F + \Delta k_F) \approx 1- (k_F + \Delta k_F)^2/2$ 
when $k_F + \Delta k_F$ is small (say $k_F + \Delta k < \pi/2$), 
one finds that the previous estimate $\propto s/a_B$ for 
$r_s^*$ is increased by an interaction effect $\propto (s/a_B)^3 $ 
for small $r_s$.

\subsection{Lattice threshold in the strong coupling limit}

 In the strong coupling limit $r_s \rightarrow \infty$, the GS energy 
is mainly classical (Coulomb energy) with weak quantum corrections.
The electron lattice minimizing the Coulomb energy can be different 
if the square lattice of the model is not commensurate 
with the hexagonal lattice that the electrons form in the continuum 
limit. This is one obvious source of lattice effects when $r_s \rightarrow 
\infty$. One does not discuss it, restricting the study to values 
of $N$ and $L$ where the two lattices are commensurate and give 
the same Coulomb energy. In this case, the lattice can nevertheless 
change the vibration modes of the electron system around its classical 
electrostatic limit, when $\hbar \rightarrow 0$. Let us consider 
the leading  quantum corrections to the classical energy.  

In the continuum model, the GS energy per particle in rydbergs 
is given by Eq. \ref{strong-coupling}. The first quantum correction 
$\Delta E_0^c (r_s \rightarrow \infty)$ to the Coulomb energy 
($\propto r_s^{-1}$) is given by the zero point energy of the harmonic 
oscillations of the electrons around their lattice points: 
\begin{equation}
\Delta E_0^c (r_s \rightarrow \infty)= \frac{f_1}{r_s^{3/2}}.
\label{continuum-correction}
\end{equation}

 In the lattice model, the classical limit $t \rightarrow 0$ is not 
described by the expansion in powers of $r_s^{1/2}$ valid 
in the continuum limit (Eq. \ref{strong-coupling}), but by a 
lattice perturbation theory where the small parameter is $t^2/U$. 
Examples of this lattice expansion valid when $t^2/U \rightarrow 0$ 
can be found in refs. \cite{wpi,ksp,np,mp,sw}. Its dominant 
quantum correction $\Delta E_0^l (r_s \rightarrow \infty)$  to the 
classical Coulomb energy comes from the term $4Nt$ of the lattice 
Hamiltonian (\ref{H-lattice-site}). 
Expressed in rydbergs ($1 Ry=U^2/4t$) per particle, this gives  
\begin{equation}
\Delta E_0^l (r_s \rightarrow \infty)= \frac{16t^2}{U^2}.  
\label{lattice-correction}
\end{equation}
The next quantum corrections of order $t^2/U$ become negligible in 
the limit $t \propto \hbar^2 \rightarrow 0$. 
 
 Let us consider a low density of electrons in a very large lattice, 
where one has the same hexagonal lattice with the same harmonic 
vibrations than in the continuum. The quantum corrections to the 
Coulomb energy and $r_s$ are then given by Eq. \ref{continuum-correction} 
and Eq. \ref{r_s-lattice} respectively. If one takes the classical limit 
$\hbar \rightarrow 0$ in such a system, $\Delta E_0^c (r_s)$ will reach 
the lattice limit  $\Delta E_0^l (r_s)$ which cannot be exceeded. This 
corresponds to coupling strengths where the harmonic vibrations of the 
electron lattice become so small that the discrete nature of the available 
space \cite{np,mp} becomes relevant. The lattice threshold $r_s^*$ 
is then obtained from the condition $\Delta E_0^c (r_s) \approx 
\Delta E_0^l (r_s)$ and the continuum approximation is only valid 
if: 
\begin{equation}
\frac{N}{L^2}<\frac{1}{\pi} \left(\frac{4^5}{f_1^4}\right)^{1/3} 
\left(\frac{t}{U}\right)^{2/3}. 
\end{equation}
Using the lattice spacing to the Bohr radius ratio $s/a_B$, 
one gets that the lattice GS can be described by a continuum theory 
in the thermodynamic limit if  
\begin{equation}
r_s > r_s^* = 0.55 \left(\frac{s}{a_B}\right)^{4/3},
\label{strong-coupling-threshold}
\end{equation}
and exhibits lattice effects otherwise. 

\subsection{Two dimensional systems of different $a_B^*$}

If the effective mass of the carriers is $m^*$ in a medium of 
dielectric constant $\varepsilon_r$, one must replace $a_B$ by the 
corresponding effective Bohr radius  $a_B^{*}$. Some  typical 
values of $m^*/m_0$, $\varepsilon_r$, $a^*_B$, $s$ and $s/a^*_B$ 
are given in Table \ref{table1} for two dimensional systems of 
charges created in various systems. In general, the lattice spacing 
$s$ is always of a few angstroms, and $\varepsilon_r \approx 10$. 
But the carriers can be light or heavy, as indicated in Table 
\ref{table1}. 

\begin{table}[t]
\begin{center}
\begin{tabular}{|c||c|c|c|c|c|} 
\hline
& $m^*/m_0$ & $\varepsilon_r$ & $a^*_B$ (\AA) & s (\AA) & $s/a^*_B$\\
\hline
\hline 
(1) & $0.19$  & $12$ & $33,2$ & $2.35$  & $0.071$ \\
\hline
(2) & $0.07 $ & $13$ & $100,0$ & $4.0$ & $0.04$ \\ 
\hline
(3) & $0.6$ & $13$ & $12 $ & $4.0$ & $0.33$ \\ 
\hline
(4) & $10 $ & $10$ & $0.53$ & $3.8$  & $7.16$ \\
\hline
(5) & $175 $ & $20$ & $0.061$  & $2.85$  & $46.7$ \\
 \hline
\end{tabular}
\end{center}
\caption{Typical physical parameters for two dimensional 
systems of charges of increasing effective masses created 
in various devices: (1) Si-Mosfet, (2) n-doped GaAs - GaAlAs 
heterostructure, (3) p-doped GaAs - GaAlAs heterostructure, 
(4) cuprate oxides exhibiting high-$T_c$ superconductivity 
and (5) layered sodium cobalt oxides Na$_x$CoO$_2$.}
\label{table1}
\end{table}

 When they are light, as in Ga-As heterostructures or in Si-Mosfets, 
$s$ is small compared to $a_B^{*}$, and $r_s^*$ is small. The immobile 
ions can be modeled by a continuum uniform jellium, unless one 
reaches the very large densities where the Fermi surface becomes 
deformed. These densities are out of reach in today's semi-conductor 
field effect devices. Therefore, if one numerically studies charge 
crystallization for those systems using a lattice model, $U/t$, $L$ 
and $N$ must be taken such that $r_s=(UL/2t)/(\sqrt{\pi N}) > 
r_s^*$. 

 In the cuprate oxides exhibiting high-$T_c$ superconductivity, the 
effective mass of the carriers is heavier, though not large enough for 
having $s/a_B^{*}$ above $23$, the value for which 
$0.55 (s/a_B^{*})^{4/3} \approx 37$. The deformation of the Fermi surface 
by the lattice yields nesting effects which can be responsible for  
singlet d-wave superconductivity \cite{ruvalds} in those oxides. The 
lattice filling is large, reaching the limit where one can take a  
Hubbard model near half-filling. When one goes away from half-filling 
by chemical doping, those systems can reach a more dilute limit where 
one gets a quantum fluid which can be described by a continuum theory and 
where the scaling parameter $r_s$ becomes relevant.    

 There are systems with much larger effective masses, as those described by 
heavy fermions theories where  $s/a_B^{*}$ can exceed $23$. One example 
is given \cite{oxyde}  by layered Lithium or Sodium Cobalt oxides:  
Li$_x$CoO$_2$ or Na$_x$CoO$_2$, where the effective mass of carriers 
can reach $\approx 200$, a value more familiar to $f$-band heavy fermions 
than to $d$-band metals, as discussed by Roger and Shannon\cite{roger}. 
Those systems look particularly interesting for the subject of this 
study, since one should observe by increasing the carrier 
concentration a continuum-lattice transition for the electron (or hole) 
crystal, before having the quantum melting of this crystal in the 
lattice solid regime, to eventually obtain at higher densities a quantum 
fluid with a deformed Fermi surface.  

\begin{figure}
\begin{center}
\epsfxsize=8cm 
\epsfbox{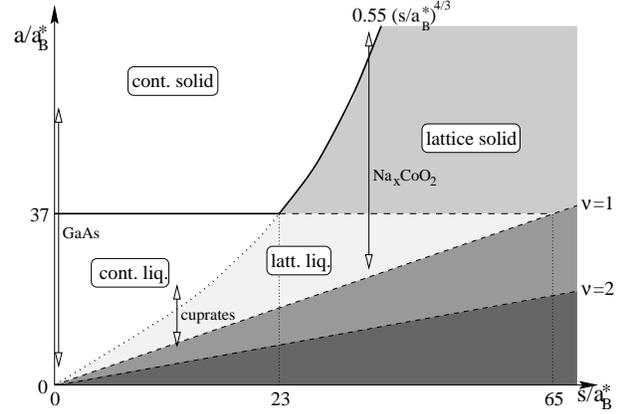}
\end{center}
\caption{Sketch of the lattice and continuum regimes in 
the $(s/a^*_B,a/a^*_B)$ plane. In the non-shaded part, 
the lattice GS can be described by a continuum theory. 
The shaded part below the dashed line $\nu=N/L^2 = 1$ 
(2 with spins) is forbidden in a lattice model (more than 1 
(2 with spins) electron per site). The line $r_s=a/a^*_B \approx 37$ 
gives the density under which Wigner crystallization is assumed 
to occur in the continuum. The thick line  $0.55 (s/a^*_B)^{4/3}$ 
gives the lattice threshold $r_s^*$ above $r_s \approx 37$, 
while the dotted line gives $r_s^* \propto s/a^*_B + O(s/a^*_B)^3$ 
below $r_s \approx 37$. The three double arrows correspond 
to typical Ga-As heterostructures, cuprate oxides and layered 
sodium cobalt oxides Na$_x$CoO$_2$. Increasing the density, one 
goes from a continuum liquid or solid towards lattice regimes.}
\label{FIG3}
\end{figure}

In Fig. \ref{FIG3}, the regime of validity of the continuum 
approximation is given in the $(s/a^*_B,a/a^*_B)$ plane, for 
typical Ga-As heterostructures, cuprate oxides or layered 
sodium cobalt oxides Na$_x$CoO$_2$. The dashed line $\nu=N/L^2 = 1$ 
corresponds to systems usually described by the half-filled Hubbard 
model, when the spins are included. Increasing the density, one 
goes towards a lattice regime where the continuum approximation 
breaks down, giving rise to a lattice liquid ($s/a^*_B<23$) 
or to a lattice solid ($s/a^*_B>23$).

\section{Scaling in a lattice model with a fixed number $N$ of 
particles}
\label{section4}

Let us define the lattice parameter suitable when $N$ is constant. 
\begin{equation}
r_l=\frac{UL}{t}=r_s (2\sqrt{\pi N}). 
\label{r_l-lattice}
\end{equation} 
So far, we have considered the realistic case where $N$ is varied 
by an electrostatic gate or by chemical doping in a lattice 
where $L$, $t$ and $U$, and hence $r_l$ are given. One can also 
study the lattice-continuum crossover by varying the lattice 
parameter $r_l$ in a system of $N$ particles. If $N$ remains 
constant in a lattice where one varies $r_l \propto r_s$, 
one gets the continuum limit and its universal scaling for 
small values of $r_l$ while the lattice becomes relevant and 
the continuum scaling breaks down for large $r_l$. 

 This can be seen from Hamiltonian (\ref{Hamiltonian-lattice-k}), 
where the components of the two dimensional vectors ${\bf k}, 
{\bf k}'$ and ${\bf q}$ can take the values $0,2\pi/L, \ldots, 2\pi(L-1)/L$.
If the Fermi energy is sufficiently small for having 
$4 - 2 (\cos k_x +  \cos k_y) \approx {\bf k}^2$ for all the 
states below the Fermi surface, the kinetic energy reads
\begin{equation}
4Nt - 2 t\sum_{ {\bf k}<{\bf k}_F} (\cos k_{ix} +  \cos k_{iy}) 
\approx {4 \pi^2 t\over L^2} \sum_{i=1}^N {\bf p}_i^2,
\end{equation}
where ${\bf k}_i = (2\pi/L){\bf p}_i$ with $p_x,p_y << L$ for 
all ${\bf k}_i<{\bf k}_F$. Expressed in rydbergs ($1Ry = U^2/4t$), 
the diagonal matrix elements of Hamiltonian (\ref{Hamiltonian-lattice-k})  
due to the kinetic energy depend only on the lattice parameter $r_l$:
\begin{equation}
{4 \pi^2 t\over L^2} \sum_{i=1}^N {\bf p}_i^2
= \frac{16 \pi^2} {r_l^{2}} \left(\sum_{i=1}^N {\bf p}_i^2\right) Ry. 
\end{equation}
The diagonal Coulomb matrix elements are given by 
$N (N-1) U V({\bf q} = 0) - \sum_{i\neq i'} UV({\bf k}_i-{\bf k}_{i'})$, 
while the off-diagonal terms $\propto U (V({\bf k}_{i_1}-{\bf k}'_{i'_1}) - 
V({\bf k}_{i_1}-{\bf k}'_{i'_2}))$, where $V({\bf q})$ is given by 
Eq. \ref{int}. If $U$ is small, only the off-diagonal 
terms with small momentum transfers (${\bf q} = (2\pi/L){\bf p}$ 
with $p_x, p_y << L$) play a role. Transforming the discrete sum to a 
continuum integral, one gets for the interaction matrix elements:  
\begin{eqnarray}
UV({\bf q}) &=& {U\over 2 L^2} \sum_{\bf j} {e^{i {\bf q j}}\over 
d_{{\bf j},0}} \nonumber \\
&=& {U\over 2 L} \sum_{\bf j} {e^{i L{\bf q} {{\bf j}\over L}}\over 
{d_{{\bf j},0}/L}} {\Delta j_x \over L} {\Delta j_y \over L} \nonumber \\
&\approx& {U\over 2 L}\int_0^1 \int_0^1 
{e^{2 \pi i{\bf p}{\bf r}}\over d({\bf r},0)} 
d^2{\bf r} = {U\over 2 L} I({\bf p}).
\end{eqnarray}
The integral $I({\bf p})$ is independent\cite{np} of $L$ 
when $p_x, p_y <<L$ and the Coulomb matrix elements of Hamiltonian 
(\ref{Hamiltonian-lattice-k}) become also a function of the lattice 
parameter $r_l$ only:
\begin{equation}
UV({\bf q})={4 t\over U^2}{U\over 2 L} I({\bf p}) Ry 
=\frac{2 I({\bf p})}{r_l} Ry. 
\end{equation}

 For $N$ fixed, assuming that $N$ is small enough 
for avoiding deformed Fermi surfaces without interaction, 
the low energy levels depend only on the lattice parameter 
$r_l$ when $r_l$ is small. The question is to determine the 
interaction threshold $r_l^*$ above which the off-diagonal 
interaction terms begin to delocalize this GS to states of 
higher momenta, where $4 - 2 (\cos k_x +  \cos k_y) \neq {\bf k}^2$. 
When $r_l$ exceeds this $r_l^*$, the lattice GS ceases to be a function 
of $r_l \propto r_s$ as in the continuum limit. 

\section{Lattice effects for a few correlated particles}
\label{section5}

We propose three criteria giving similar lattice thresholds $r_l^*$ 
for an interacting system of $N$ polarized electrons, which 
will be carefully studied when $N=3$ in the next section. 
The first one is a delocalization criterion in the Fock 
basis of lattice site orbitals. The second one uses the 
invariance of the persistent current when one varies the 
interaction strength in the continuum, an invariance which can be 
broken by the lattice. The third one is based on the limit 
imposed by the lattice to the zero point energy of the 
harmonic vibrations of an $N$ electron solid, as previously 
discussed in the thermodynamic limit.  

\subsection{Criterion 1: Delocalization in the Fock basis of 
lattice site orbitals}
\label{subsection5.1}

 Let us  consider the system of $N$ particles in real space instead 
of reciprocal space, in the limit $t=0$ where the $N$ electrons are 
localized on $N$ sites (see Fig. \ref{Fig2}) and form states 
$\left|J\right>=c_{\bf{j}_1}^{\dagger} \ldots c_{\bf{j}_N}^{\dagger} 
\left| 0 \right>$ of energy $E_{Coul}(J)$. As one turns on $t$, one 
can expect that the lattice becomes irrelevant as each electron ceases to be 
localized on a single site. In analogy with the problem of a single 
particle in a disordered lattice, one can use the criterion first 
proposed by Anderson \cite{anderson}: delocalization takes place when 
the hopping term $t$ between directly coupled sites becomes of the 
order of their energy spacing $\Delta E$. This criterion was extended 
to interacting systems in many different contexts: onset of quantum chaos 
in many body spectra \cite{wpi,wp,shepelyansky-suskhov} and in the quantum 
computer core \cite{benenti}, quasi-particle lifetime and delocalization 
in Fock space \cite{altshuler,jacquod}. In our case, the states 
become delocalized in the many body basis built from the states 
$\left|J\right>$ when the matrix element $\left<J'|H_{kin}|J\right>$ of 
the one body perturbation $H_{kin} \propto t$ coupling a state 
$\left|J\right>$ to the ``first generation'' of states $\left|J'\right>$ 
directly coupled to it by $H_{kin}$ exceeds their energy spacing 
$\Delta E_{Coul} = E_{Coul}(J')-E_{Coul}(J)$. 
This gives $t > \Delta E_{Coul}$. Applying this criterion to 
the GS, one obtains $r_l^*$ from the condition 
\begin{equation}
t \approx \Delta E_{Coul} ,   
\label{Anderson}
\end{equation}     
where $\Delta E_{Coul}$ is the increase of Coulomb energy yielded 
by the hop of one particle localized on the GS configuration to a nearest 
neighbor site when $t=0$. When $t$ exceeds $\Delta E_{Coul}$, 
the GS is delocalized on the $J$-basis, and hence on the lattice, and 
the lattice GS behaves as the continuum GS.

\subsection{Criterion 2: Persistent currents}
\label{subsection5.2}

  Since a continuum model is invariant under translations, the motion 
of the center of mass can be decoupled from the relative motions. 
Thus the continuum Hamiltonian $H_c$ (Eq. \ref{H-continuous}) can be 
decomposed in two parts, one related to the center of mass motion which 
is independent of the interaction, while the second one contains only 
the relative motions and hence the interaction. 
This has a very important consequence for the persistent current $I$ 
driven by an enclosed Aharonov-Bohm flux $\Phi$ in a continuum model: $I$ 
is independent of $r_l$ and keeps its non interacting value. For having 
the topology of a $2d$ torus enclosing $\phi$ along the $x$-direction, 
one takes the corresponding curled BC in this direction, keeping periodic 
BC in the $y$-direction. For a sufficient $r_l$, the electrons form a 
Wigner solid and the small relative motions cannot feel the BCs. In this 
limit, $I$ is just given by the center of mass motion, which is independent 
of $r_l$, and hence coincides with its non-interacting value. This point 
remains correct for small $r_l$, as it was proven for $1d$-rings 
\cite{muller-groeling,krive,burmeister} and observed for $d=2$ 
\cite{muller-groeling}. In contrast, since the previous decomposition 
into two parts does not necessary hold for $H_l$, $I \neq I(r_l=0)$ 
for a lattice when  $H_l$ and $H_c$ have different GSs. The decay of $I$ 
above $r_l^*$ (small $t/U$ at fixed $N$ and $L$) can be evaluated 
\cite{ksp,np,sw} by the leading contribution (of order $N$) $I_{l}^{(N)}
\propto t(t/U)^{N-1}$ of the $t/U$ lattice expansion. The value of $r_l$ 
for which
\begin{equation}
I(U=0) \approx I_{l}^{(N)} 
\end{equation}
gives the Criterion 2 for $r_l^*$ 
(see Fig. \ref{Fig2} right). Instead of $I(\Phi)$, one can prefer 
to use the Kohn curvature $C_K=\partial^2 E_0/\partial \Phi^2$ evaluated 
at $\Phi=0$ or the GS energy change $\Delta E_0= 
E_0(\Phi=0)-E_0(\Phi=\Phi_0/2)$ where $\Phi_0$ is the flux quantum. 
To apply $\Phi_0/2$ corresponds to have anti-periodic BC in the $x$-direction.

\subsection{Criterion 3: Lattice limit for the zero point motion of 
an electron solid}
\label{subsection5.3}

This is the criterion that we have already used in the qualitative 
discussion of the thermodynamic limit, and which we consider when $N$ 
is finite.  When $t/U \rightarrow 0$, the leading correction to the 
Coulomb energy of $H_l$ is $4Nt$. Since the correction $E_{vib} (r_l)$ 
to the Coulomb energy coming from the zero point vibrational motion of 
the continuum solid cannot exceed this lattice limit $4Nt$, $r_l^*$ 
can be obtained from the condition 
\begin{equation}
E_{vib} (r_l^*) \approx 4Nt,
\label{Vibration} 
\end{equation}  
assuming that the values of the lattice parameters can yield a Wigner 
solid for $r_l < r_l^*$. 

\section{Numerical study of three polarized electrons}
\label{section6}
 
 When one takes periodic BCs, a convention has to be chosen for 
the distance $r$ (Hamiltonian (\ref{H-continuous})) or 
$d_{\bf jj'}$ (Hamiltonian (\ref{H-lattice-site})). 
For a finite square with periodic BCs, one possible definition 
is given by:
\begin{equation}
d_{\bf jj'}^{PSC}=\sqrt{\min(|d_{x},L-|d_{x}|)^2+ \min(|d_{y}|,L-|d_{y}|)^2}, 
\label{distance1}
\end{equation}
where $d_{x}=j_x-j'_x$ and $d_{y}=j_y-j'_y$.
Hereafter, we refer to the corresponding $1/|d_{\bf jj'}|$ repulsion 
as the periodic singular Coulomb (PSC) repulsion, since it has a cusp 
when the interparticle distance $d_{\bf jj'}$ has one of its coordinates 
equal to $L/2$.  This cusp being unphysical, we introduce the 
periodic regularized Coulomb (PRC) repulsion, defined from
\begin{equation}
d_{\bf jj'}^{PRC}=\frac{L}{\pi} \sqrt{\sin^2\frac{|d_{x}|\pi}{L}+
\sin^2\frac{|d_{y}|\pi}{L}}
\label{distance2}
\end{equation}
which locally coincides with the PSC repulsion, but remains analytic 
for all values of $d_{\bf jj'}$ when $s \rightarrow 0$. The PRC repulsion 
is essentially equivalent to the Ewald repulsion obtained from the 
periodic repetition of the considered system. 

Defining $d_{\bf jj'}$ with Eq. \ref{distance2}, we calculate the 
quantities used in the different criteria for $N=3$ polarized 
electrons on a square lattice. We give in 
Appendix \ref{Appendix C} the same analysis defining $d_{\bf jj'}$ 
with Eq. \ref{distance1} instead of Eq. \ref{distance2}. The choice 
of the PRC or PSC repulsions, or of the repulsion obtained after 
Ewald summation is arbitrary for three particles in a toroidal geometry.  
Nevertheless, it allows us to check if our three proposed criteria 
for $r_l^*$ give consistent results when the long range form of the 
Coulomb repulsion is changed. The presented results extend to larger 
$L$ previous studies of the case $N=2$ and $N=3$ given in Ref.\cite{mp}
and Ref.\cite{np} respectively. 

\begin{figure}
\begin{center}
\epsfxsize=8cm 
\epsfbox{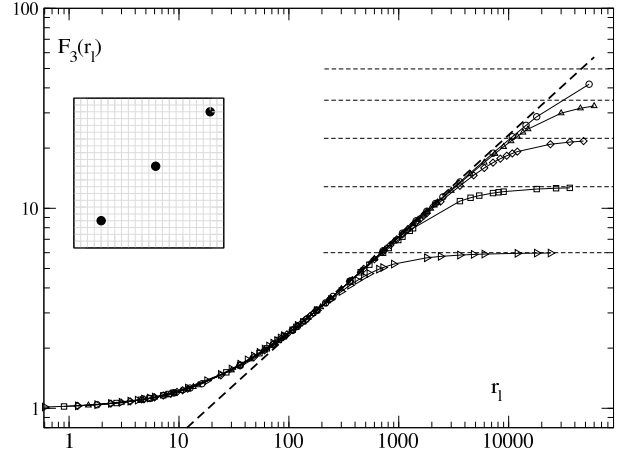}
\end{center}
\caption
{Energy ratio $F_{N=3}(L,U,t)$ as a function of $r_l=UL/t$ given by 
the PRC repulsion for $L=6$ ($\triangleright$), 9 ($\Box$), 12 
($\diamond$), 15 ($\triangle$), 18 ($\circ$). The dotted-dashed line 
gives the behavior $0.2327 \sqrt {r_l}$ (harmonic vibrations of the 
continuum Wigner molecule) and intersects the limiting dashed lines 
$12t/(4t-4t\cos(2\pi/L))$ at the $r_l^*(L)$ corresponding to 
criterion 3. Inset: A GS configuration when $t=0$ and $L=24$.}
\label{Fig4}
\end{figure}

For $t=0$, the configuration of particles minimizing the PRC Coulomb 
energies is given in the inset of Fig. \ref{Fig4}. The values of 
$L=6,9,12,15,18,\ldots$ yield a diagonal Wigner molecule shown in 
the inset which is commensurate with the square lattice. Moving one 
particle by a single hop increases the Coulomb energy $E_0=({\sqrt 6} 
U \pi)/L$ by an amount 
\begin{equation}
\Delta E_{Coul} \approx  
\frac{7\sqrt{2}\pi^{3}U}{12\sqrt{3}L^3} \ \ 
\label{prc}
\end{equation} 
when $L$ is sufficiently large.

For $U=0$, the GS energy is given by $E_0(0)=12t-8t-4t\cos(2\pi/L)$ 
for periodic BCs and becomes 
$E_0(\Phi_0/2)=12t-8t\cos(\pi/L) -4t\cos(3\pi/L)$ 
when one twists the BC in the $x$-direction. The difference 
$\Delta E_0=E_0(\Phi_0/2)-E_0(0)\approx -14 \pi^2 t/L^2$ when 
$L \rightarrow \infty$. When $t/U$ is small, $\Delta E_0$ can be 
calculated at the leading order of a $t/U$-expansion \cite{np} for $N=3$. 
This gives when $L$ is large: 
\begin{equation}
\lim_{r_l \rightarrow 0} \Delta E_0 \approx \frac{14 \pi^2 t}{L^2} 
\ \ ;\ \ 
\lim_{r_l \rightarrow \infty} \Delta E_0 \approx \frac{9 \pi^2 
t^3}{L^2 \Delta E_{Coul}}
\label{current}
\end{equation}
where  $\Delta E_{Coul}$ is given by the Eq. \ref{prc}. 
Using these expressions, one obtains from the two first criteria:
\begin{equation}
r_l^*(L) = A L^{\alpha}
\label{lattice-threshold}
\label{threshold}
\end{equation}
where $\alpha =4$ for the PRC repulsion, the constant $A$ slightly 
depending on the taken criterion.

\subsection{Persistent currents}

 We now present numerical results obtained using the Lanczos 
algorithm, using Hamiltonian (\ref{Hamiltonian-lattice-k}) and 
considering the sub-space of total momentum ${\bf K}=0$ \cite{np} 
for periodic BCs (no applied flux). 

\begin{figure}
\begin{center}
\epsfxsize=8cm 
\epsfbox{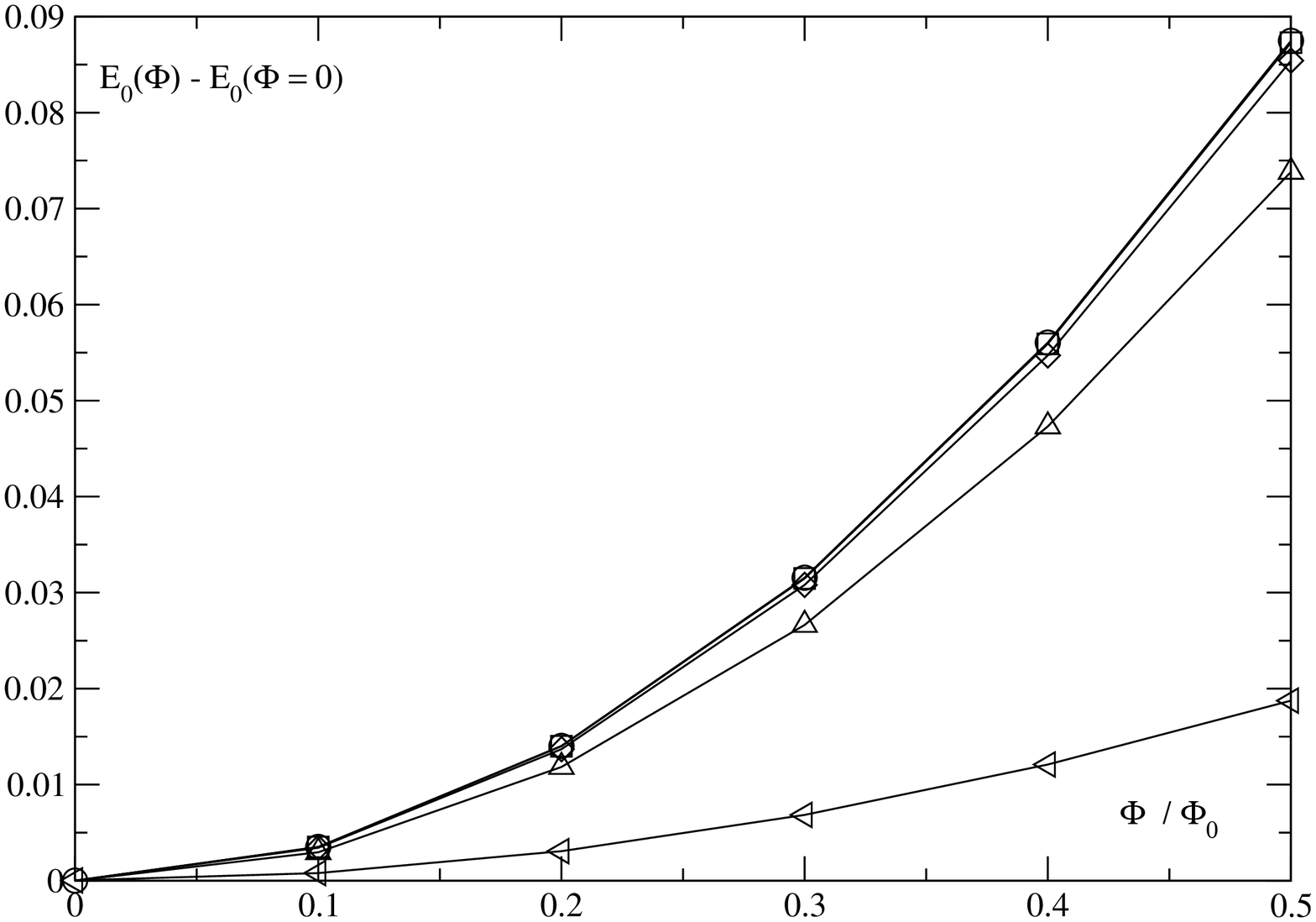}
\end{center}
\caption{
GS energy $E_0(\Phi)-E_0(\Phi=0)$ as a function of the enclosed 
dimensionless magnetic flux $\Phi/\Phi_0$ for $N=3$, $L=18$, PRC 
interaction at $r_l=6$ ($\circ$), 60 ($\Box$), 600 ($\diamond$) 
6000 ($\triangle$) and 60000 ($\triangleleft$).}
\label{Fig5}
\end{figure}

\begin{figure}
\begin{center}
\epsfxsize=8cm 
\epsfbox{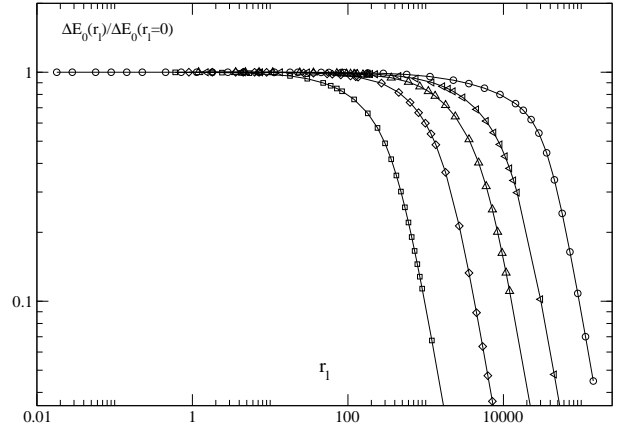}
\end{center}
\caption
{Dimensionless change $\Delta E_0 (r_l)/ \Delta E_0 (r_l=0)$ of the 
GS energy when the longitudinal BC is twisted for $L=6$ ($\Box$), 
9 ($\diamond$) ,12 ($\triangle$), 15 ($\triangleleft$), 18 ($\circ$) 
and  $N=3$ as a function of $r_l$ (PRC repulsion). 
}
\label{Fig6}
\end{figure}

Our system has the topology of a $2d$ torus. To enclose an 
Aharonov-Bohm flux $\Phi$ along the $x$-direction, one takes 
the corresponding curled BC in this direction while the BC in 
the $y$-direction remains periodic. To apply half a flux quantum 
($\Phi=\Phi_0/2$) is equivalent to take anti-periodic BC along 
the $x$-direction. In Fig.\ref{Fig5}, the increase $E_0(\Phi)-E_0(0)$ 
of the GS energy $E_0$ is given as a function of $\Phi/\Phi_0$ for 
different values of $r_l$  using a $18 \times 18$ 
square lattice. When $r_l$ is small, the curves coincide. This is 
the continuum regime where the persistent current is independent 
of the interaction. When $r_l$ is large, the increase $E_0(\Phi)-E_0(0)$ 
becomes weaker. This is the lattice regime where the persistent current 
decays as the interaction increases. One gives in Fig.\ref{Fig6} 
the dimensionless change $\Delta E_0 (r_l)/ \Delta E_0 (r_l=0)$ of 
the GS energy when the BC is twisted in the $x$-direction for 
increasing values of $L$. One can see the two limits given by 
Eq. \ref{current}, $\Delta E_0 (r_l)/ \Delta E_0 (r_l=0)\approx 1$ 
in the continuum limit, followed by a decay when $r_l$ exceeds the 
lattice threshold $r_l^*$.

\subsection{Harmonic vibrations of the continuum Wigner molecule}

For the third criterion, one needs the zero point vibrational 
energy of the Wigner molecule that the three particles form 
when $r_l$ is large, but smaller than $r_l^*$. This can be 
calculated in the continuum limit, using for $N=3$ the same 
expansion in powers of $r_s^{1/2}$ than those used in 
Eq. \ref{strong-coupling} for $N \rightarrow \infty$. We summarize 
the main points, the details being given in Appendix 
\ref{Appendix B}. In the continuum, the Hamiltonian $H_c$ can 
written as the sum of two decoupled terms. Denoting 
${\bf R}=(\sum_i^3 {\bf r}_i)/3$ the coordinate of the 
center of mass, the first term reads $H_{CM}=(\hbar^2/6m) \nabla^2_{\bf R}$ 
and corresponds to the rigid translation of the molecule while the other 
term contains the relative motions and the interaction. For a Wigner 
molecule, the second part can be simplified and expressed in terms of 
the normal coordinates suitable for describing the small vibrations 
around equilibrium. 

 The PRC repulsion is harmonic around equilibrium, and the three particles 
form a diagonal chain as indicated in the inset of Fig. \ref{Fig4}
when $L/3$ is integer. One gets four decoupled harmonic oscillators, 
two corresponding to a longitudinal mode of frequency $\omega_l=\sqrt{20 B}$, 
the two others being a transverse mode of frequency $\omega_t=\sqrt{8B}$, 
where $B=(\sqrt{6} e^2 \pi)/(24 D^3 m)$. The zero point vibrational energy 
is then given by: 
\begin{eqnarray}
E_{vib} (r_s,N=3)
&=& \hbar (\omega_l+\omega_t) 
\nonumber \\
&=& 2\pi \frac{\sqrt{5} +\sqrt{2}}{\sqrt{18}} 
\left(\frac{2}{\pi} \right)^{1/4} r_s^{-\beta}
\label{Vibration-3-PRC}
\end{eqnarray} 
in rydbergs where $\beta=3/2$, with $r_s=r_l/(2\sqrt{3\pi})$ for $N=3$.

\subsection{Scaling of the ground state energy}

From the GS energy $E_0(L,U,t)$ of ${\bf K}=0$, and for a given value 
of $N$, we define the dimensionless ratio $F_N(L,U,t)$ by: 
\begin{equation} 
F_N(L,U,t)=\frac{E_0(L,U,t)-E_0(L,U,t=0)}{E_0(L,U=0,t)}.
\label{ratio}
\end{equation} 
This ratio gives the change of the GS energy from the Coulomb 
energy due to the quantum effects, divided by the GS energy 
without interaction. 

The results for the PRC repulsion are shown in Fig. \ref{Fig4}. 
For $t=0$, the values of $L=6,9,12,15,18$ are 
commensurate with the period of the diagonal Wigner molecule 
shown in the inset. This gives the same classical Coulomb 
energy for the lattice and the continuum when $t\rightarrow 0$, 
eliminating a trivial source of lattice effects. 
When $F_{N=3}(L,U,t)$ is plotted as a function of $r_l$,  
the different functions $F_{N=3}(L,U,t)$ scale without an 
observable lattice effect up to the $r_l^*(L)$ 
exactly given by Criterion 3. Using $E_0(L,U=0,t)=12t-8t-4t\cos(2\pi/L)$ 
one can see that the numerical results coincide with the analytical 
result $F_{N=3} = 0.2327 \sqrt{r_l}$ implied by Eq. \ref{Vibration-3-PRC} 
for intermediate values of $r_l$ where one has a continuum Wigner molecule. 
The function  $F_{N=3}(L,U,t)$ saturates to $4Nt/E_0(L,U=0,t)$ above 
$r_l^*(L)$, as indicated by the dashed lines.

\section{Effect on the scaling function when $N$ varies}
\label{section7}

\begin{figure}
\begin{center}
\epsfxsize=8cm 
\epsfbox{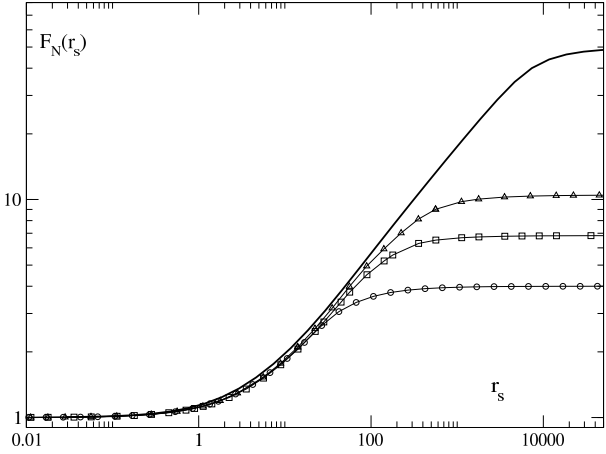}
\end{center}
\caption
{Energy ratios $F_{N} (L,U,t)$ using the PRC repulsion 
for $N=3$ ($L=18$ solid line) and $N=4$  with $L=6$ ($\circ$), 8 
($\Box$), 10 ($\triangle$) as a function of $r_s=r_l/(2\sqrt{N\pi})$.  
(PRC repulsion). 
}
\label{Fig7}
\end{figure}

In Fig. \ref{Fig7}, a small change of the scaling curve 
$F_{N}(r_s=r_l/(2\sqrt{N\pi})$ can be seen when a fourth electron is 
added, accompanied by the expected breakdown of the  scaling behavior 
above the corresponding $r_l^*$ for $N=4$. When $N \rightarrow \infty$, 
$F_{N}$ should converge towards a thermodynamic limit depending only 
on $r_s$. Unfortunately, a study of this convergence is out of reach 
of a numerical approach using exact diagonalization.

\section{Conclusion}
\label{section8}

 We have studied the lattice effects upon an interacting system 
of polarized electrons in two dimensions. We have first considered 
the case where the number $N$ of polarized electrons is increased 
in a square lattice of large size $L$ and of fixed parameters 
$U=e^2/(\varepsilon_r s)$ and $t=\hbar^2/(2m^*s^2)$. This corresponds 
to semi-conductor field effect devices or layered oxides where the 
number of carriers can be varied by an electrostatic gate or by chemical 
doping. Starting from an empty lattice, one has a continuum regime and its 
universal scaling if one uses the parameter $r_s$, until the carrier 
density $n_s^*=1/(\sqrt {\pi} r_s^* a^*_B)^2$ is reached. At this density, 
the continuum approximation with its universal scaling breaks down, as 
sketched in the phase diagram given in Fig. \ref{FIG3}. This lattice 
threshold $r_s^*$ takes place in a quantum fluid phase if the 
carriers are light and the ratio $s/a^*_B < 23$. We have studied more 
particularly in the remaining part of our manuscript the case of heavy 
carriers where $s/a^*_B > 23$, for which the continuum approximation 
breaks down in the crystalline phase. We have pointed out 
the limit imposed by the lattice of positive ions upon the zero 
point motion of the electron lattice. We have neglected the obvious 
problem coming from the incommensurability of the two lattices, 
and focus our attention to the commensurate case. The studied lattice 
effects are independent of this commensurability issue, which should 
matter at large lattice fillings. 

 In the second part of this manuscript, we have studied the role of 
a lattice for a fixed number $N=3,4$ of polarized electrons. The 
lattice-continuum crossover is then obtained by varying the lattice 
parameter $r_l=(UL)/t$. $r_l$ and $r_s$ play the same role when $N$ 
is fixed, since $r_s=r_l/(2\sqrt{\pi N})$. The continuum approximation 
is valid and there is a universal scaling when one uses the parameter 
$r_l$ as far as $r_l$ does not exceed a lattice threshold $r_l^*$ 
which has been determined from three criteria. One of them was based 
on the behavior of the persistent current $I(r_l)$ driven by an enclosed 
Aharonov-Bohm flux. For $r_l < r_l^*(L)$, $I(r_l)=I(r_l=0)$ while 
$I(r_l)$ decays above $r_l^*$ and $r_l$ ceases to be a scaling 
parameter. For a finite number of particles, one goes from the continuum 
regime towards the lattice regime through a smooth crossover. 
The question to know if this smooth crossover does not become 
sharper when $N \rightarrow \infty$, to give rise to a true quantum 
transition is an interesting issue which we postpone to a following 
study.

\begin{acknowledgement}
Z. \'A. N\'emeth acknowledges the financial support provided 
through the European Community's Human Potential Programme under 
contract HPRN-CT-2000-00144 and the Hungarian Science Foundation 
OTKA T034832.
\end{acknowledgement} 

\appendix
\section{Weak interaction correction to $r_s^*$}
\label{Appendix A} 

When $r_s$ is small, the main effect of the interaction is to smear 
the Fermi surface, giving an uncertainty $\Delta k_F$ to $k_F$, 
such that one expects to have the continuum behavior when 
$k_F + \Delta k_F$, and not only $k_F$, is small. To evaluate 
$\Delta k_F$, we assume that the low excited states only become 
occupied at low $r_s$. The first excitation energy reads: 
\begin{eqnarray}
\Delta E_{1,0} &=& t \sum_{i=1}^N ({\bf k}^2_{i1} - {\bf k}^2_{i0}) 
\nonumber \\
&=& 2 t ({\bf k}'^2-k_F^2) \approx 4t k_F \Delta k
\end{eqnarray}
where the factor 2 comes from momentum conservation and 
$k'$ is the wave vector of an empty state above the 
non interacting Fermi surface. This gives us the relation:
\begin{equation}
{\Delta E_F\over \Delta k_F} \approx 4 t k_F.
\end{equation}
The Fermi energy uncertainty $\Delta E_F$ can be estimated from 
the spreading $\Gamma$ of a non interacting level, when one turns 
on the interaction. Using Fermi's golden rule, one gets 
\begin{equation}
\Gamma \approx {2\pi |H_{0,1}|^2 n(k_F) \over\Delta E_{1,0}}, 
\end{equation}
where the matrix element of interaction coupling the GS to the first 
excited state $|H_{0,1}| = 2U (V({\bf q})-V(2k_F)) \approx 2 U V({\bf q})$ 
reads:
\begin{equation}
H_{01} \approx {U\over L^2} \sum_{\bf j} {e^{i{\bf qj}}\over d_{\bf j0}} 
 \approx {U\over L} \int_0^1 \int_0^1 {\cos{2\pi x}\over d({\bf r},0)} d^2{\bf r} = \frac{I U}{L}.
\end{equation}
$I$ is a constant equal to  $1.029$ for the PRC repulsion. 
The number of states $n(k_F)$ on the Fermi surface is equal to  
$\sqrt{4 N \pi}$ and $\Delta E_{10}=4t k_F |{\bf q}|$ where 
$|{\bf q}|=(2\pi/L)$ is the smallest momentum for an excitation. 

One eventually gets for the Fermi energy uncertainty
\begin{equation}
\Delta E_F \approx \Gamma \approx {U^2\over 4 t} I^2,
\end{equation}
which gives for the Fermi momentum uncertainty
\begin{equation}
\Delta k_F \approx \left({UI \over t}\right)^2  {L\over 32 \sqrt{\pi N}}.
\end{equation}
The condition  
\begin{equation}
k_F + \Delta k_F < {\pi \over 2}
\label{cond}
\end{equation}
is satisfied if   
\begin{equation}
r_s - {I^2 \over 16}r_s^3 > {4\over \pi} {s\over a_B}
\end{equation}
When $r_s$ is not too large, the continuum theory is valid 
if $r_s> r_s^*$ with a threshold $r_s^*$ having a small 
correction $\propto (s/a_B)^3$ driven by the interaction:  
\begin{equation}
r_s^* \approx {4\over \pi} {s\over a_B} + {4 I^2 \over \pi^3} 
\left({s\over a_B}\right)^3. 
\end{equation}
 The constants in the expression of $r_s^*$ depend on the used 
criterions for neglecting lattice effects (for instance $4/ \pi$ 
comes from the condition (\ref{cond})).

\section{Zero point energy of a continuum Wigner molecule 
for $N=3$}
\label{Appendix B} 

 For three spinless fermions on a continuum square domain of size $D$ 
with periodic BCs, the continuum PRC repulsion reads 
\begin{equation}
V({\bf r})= {e^2 \pi\over D 
\sqrt{\sin^2 {r_x \pi\over D}+
\sin^2 {r_y \pi \over D}}}.
\label{sinus-potential}
\end{equation}
If $D$ is large enough, the GS is a "Wigner molecule" of delocalized 
center of mass, but of quasi-localized inter-particle spacings for 
minimizing the Coulomb energy. For a certain center of mass, 
the molecule of lowest Coulomb energy with the repulsion 
(\ref{sinus-potential}) consists in putting the particle coordinates 
at ${\bf r}_1=(0,0)$,  ${\bf r}_2=(D/3,D/3)$ and ${\bf r}_3=(-D/3,-D/3)$. 
This configuration has the Coulomb energy 
\begin{equation}
E_{Coul} = \sqrt{6} e^2 \pi/D.
\end{equation}
The particles forming this molecule vibrate around the equilibrium 
positions. This motion is an harmonic oscillation if the amplitude
of the vibration is small. To describe this harmonic motion, one 
expands the pair-potential (\ref{sinus-potential}) around the equilibrium 
distance ${\bf r}_0=(D/3,D/3)$ up to the second order:
\begin{eqnarray}
V({\bf r}) &\approx& {\sqrt{6}e^2\pi\over3D}+ (\dots) \nonumber \\
&+& {7\sqrt{6}\over 72}{e^2 \pi\over D^3} \left(\left(x-{D\over 3}\right)^2+
\left(y-{D\over 3}\right)^2\right)  \nonumber \\
&+&{\sqrt{6}\over 12}{e^2 \pi\over D^3}
\left(r_x-{D\over 3}\right) \left(r_y-{D\over 3}\right)+ O(r^3), 
\label{sinus-expansion}
\end{eqnarray}
where the missing term $(\dots)$ is the first order 
contribution which disappears after summing over all the pair 
potentials. The expansion (\ref{sinus-expansion}) becomes:
\begin{equation}
V({\bf r})\approx {\sqrt{6}e^2\pi\over3D}+ (\dots) + ({\bf r}-{\bf r}_0)
\left(\matrix{A & B \cr B & A}\right) 
({\bf r}-{\bf r}_0).
\end{equation}
where 
\begin{equation}
A = {7\sqrt{6}\over 72}{e^2 \pi\over D^3} 
\end{equation}
and $B = 3A/7$. 

The three particle Hamiltonian with the expanded repulsion 
becomes $H_c \approx E_{Coul}+H_{harm}$, where the harmonic 
part is:
\begin{equation}
H_{harm} = -{\hbar^2 \over 2 m}(\nabla_1^2+\nabla_2^2+\nabla_3^2)+ 
{\vec X} {\hat M} {\vec X}.
\label{Hamilton-expansion}
\end{equation}
The vector ${\vec X}=(x_1,y_1,x_2,y_2,x_3,y_3)$ is composed 
of the 6 relative coordinates and the $6\times 6$ matrix ${\hat M}$ 
is given by:
\begin{equation}
{\hat M} = \left(\matrix{2A&2B&-A&-B&-A&-B\cr
2B&2A&-B&-A&-B&-A\cr -A&-B&2A&2B&-A&-B\cr 
-B&-A&2B&2A&-B&-A\cr -A&-B&-A&-B&2A&2B\cr
-B&-A&-B&-A&2B&2A}\right).
\end{equation}

Diagonalizing ${\hat M}$, one obtains the normal modes 
of the harmonic oscillations while the eigenvalues of 
${\hat M}$ give their frequencies. One obtains
\begin{itemize}
\item Two eigenvectors of eigenvalue $0$.
\begin{eqnarray}
{\vec \chi_1}&=&{1\over\sqrt{3}}(1,0,1,0,1,0) \cdot {\vec X}, 
\nonumber \\
{\vec \chi_2}&=& {1\over\sqrt{3}}(0,1,0,1,0,1)\cdot{\vec X}.
\end{eqnarray}
This zero frequency mode corresponds to the translation of the center 
of mass of the molecule. 

\item Two other eigenvectors of eigenvalue $10B$, 
corresponding to the longitudinal mode (vibration parallel 
to the axis of the molecule). The normal coordinates can be 
taken as:
\begin{eqnarray}
{\vec \chi_3}&=& {1\over2}(1,1,-1,-1,0,0)\cdot{\vec X}, \nonumber \\
{\vec \chi_4}&=& {1\over\sqrt{12}}(1,1,1,1,-2,-2)\cdot{\vec X};
\end{eqnarray}

\item Two eigenvectors of eigenvalue $4B$, corresponding to the transverse 
modes. The normal coordinates can be taken as:
\begin{eqnarray}
{\vec \chi_5}&=& {1\over2}(1,-1,-1,1,0,0)\cdot{\vec X}, \nonumber \\
{\vec \chi_6}&=& {1\over\sqrt{12}}(1,-1,1,-1,-2,2)\cdot{\vec X}.
\end{eqnarray}
\end{itemize}

Using these normal coordinates, the Hamiltonian (\ref{Hamilton-expansion}) 
becomes a decoupled sum of two harmonic oscillators:
\begin{equation}
H_{harm} = -{\hbar^2 \over 2 m}\sum_{\alpha=1}^6 {\partial^2\over 
\partial\chi_\alpha^2} + 10B (\chi_3^2+\chi_4^2) + 4B 
(\chi_5^2+\chi_6^2), 
\end{equation}
For a GS of total momentum ${\bf K}=0$, there is no motion of the 
center of mass, the GS wave-function does not depend on $\chi_1$ 
and $\chi_2$ and can be factorized as:
\begin{equation}
\Psi(\chi_1,\dots,\chi_6) =  \varphi_{0L}(\chi_3) \varphi_{0L}(\chi_4)
\varphi_{0T}(\chi_5) \varphi_{0T}(\chi_6)
\end{equation}
where $L,T$ refers to the transverse and longitudinal modes
and $\varphi_0$ to the ground state of an harmonic oscillator:
\begin{equation}
\varphi_0(x)= {1\over l_\omega^{1/2} \pi^{1/4}} 
\exp-\frac{x^2}{ 2l_\omega^2}, 
\end{equation}
of length $l_\omega = \left(\hbar^2/(m^2 \omega^2)\right)^{1/4}$. 
One eventually obtains for the GS energy with the expanded 
pair potentials:
\begin{eqnarray}
E_0 - E_{Coul} &=& \hbar (\omega_T+\omega_L); \\ 
\omega_L &=& \sqrt{20 B\over m}, \\
\omega_T &=& \sqrt{8 B\over m}
\end{eqnarray}
and using the expression of $B$:
\begin{eqnarray}
E_0 - E_{Coul} &=& \sqrt{{20 \sqrt{6}\over 24}{\hbar^2e^2\pi^3\over D^3 m}} 
+\sqrt{{8 \sqrt{6}\over 24}{\hbar^2e^2\pi^3\over D^3 m}} \nonumber \\
&=& (\sqrt{5}+\sqrt{2}) \sqrt{\sqrt{6}\pi^3\over 3} \sqrt{Ut \over L^3}.
\end{eqnarray}

 For the energy ratio $F_{N=3}(L,U,t)$, using for the kinetic energy 
in the continuum limit $E_0(L,U=0,t) = 8\pi^2 t/L^2$, one gets the 
behavior numerically obtained from the lattice Hamiltonian $H_l$ and 
shown in Fig. \ref{Fig4} for intermediate $r_s$:
\begin{eqnarray}
F_0(r_s) &=& {\sqrt{5}+\sqrt{2}\over 8 \pi^2} 
\sqrt{\sqrt{6}\pi^3\over 3} \sqrt{UL\over t} \nonumber \\
&=& {\sqrt{5}+\sqrt{2}\over \sqrt{96}} 
\left({18\over \pi}\right)^{1/4} \sqrt{r_s} \nonumber \\
&=& 0.5764 \sqrt{r_s} = 0.2327 \sqrt{r_l}.
\end{eqnarray}

\section{Lattice threshold $r_l^*$ using the PSC repulsion}
\label{Appendix C}

\begin{figure}
\begin{center}
\epsfxsize=8cm 
\epsfbox{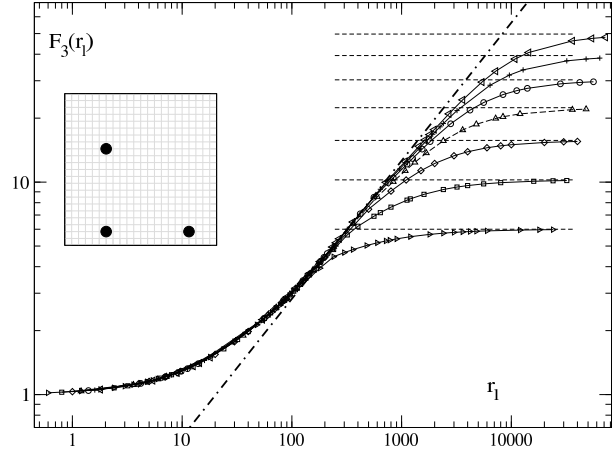}
\end{center}
\caption{
Energy ratio $F_{N=3}(L,U,t)$ as a function of $r_l=UL/t$ using the 
PSC repulsion for $L=6$ ($\triangleright$), 8 ($\Box$), 10 
($\diamond$), 12 ($\triangle$), 14 ($\circ$), 16 ($+$), 18 ($\triangleleft$). 
The dotted-dashed line gives the $r_l^{2/3}$ behavior due to the vibrations 
of the continuum Wigner molecule. Inset: a GS configuration when $t=0$ 
and $L=24$.
}
\label{Fig8}
\end{figure}

 With the distance $d_{\bf jj'}$ defined by Eq. \ref{distance2}, 
we have previously studied the validity of a continuum approximation for 
a lattice model of three polarized electrons. In this appendix, 
we revisit the same issue defining $d_{\bf jj'}$ from Eq. \ref{distance1} 
instead of Eq. \ref{distance2}. Let us calculate the quantities used 
for the three criteria when one uses the PSC repulsion. For $t=0$, 
the Wigner ``molecule'' minimizing the PSC Coulomb energy has the 
triangular shape shown in the inset of Fig. \ref{Fig8}, instead of 
the linear shape shown in the inset of Fig. \ref{Fig4}. Moving one 
particle by a single hop in this triangular molecule increases the 
PSC Coulomb energy by an amount 
\begin{equation}
\Delta E_{Coul}^{(PSC)} \approx  \frac{\sqrt{2}U}{L^2} 
\label{psc}
\end{equation}
when $L$ is sufficiently large, instead of the 
$\Delta E_{Coul}^{(PRC)} \propto U/L^3$ given by Eq. \ref{prc}. 
 
For the energy change $\Delta E_0$, one obtains the same expressions 
as in Eq. \ref{current}, but with $\Delta E_{Coul}$ given by the Eq. 
\ref{psc} instead of  Eq. \ref{prc}. The two first criteria 
gives 
\begin{equation}
r_l^*(L) = A L^{\alpha},
\end{equation}
where $\alpha=3$ for the PSC repulsion, instead of $\alpha =4$ for the 
PRC repulsion. 

 When one takes the PSC repulsion, the three relative distances 
at equilibrium are precisely ${\bf r}=(L/2,L/2)$, ${\bf r}=(0,L/2)$ and 
${\bf r}=(L/2,0)$ respectively when $L$ is even. The potentials 
$v(\delta{\bf r})$ felt by the electrons around their equilibrium positions 
are singular and instead of the analytical expansion 
(\ref{sinus-expansion}) of $v(\delta{\bf r})$, one has 
$v(\delta{\bf r}) \approx C_1 |\delta r_x|+ C_2|\delta r_y|$, where 
$C_1$ and $C_2$ depend on the equilibrium positions and are 
$\propto e^2/D^2=U/L^2$. For a single particle in a $1d$-potential 
$v(x)=C|x|$, the GS energy $\epsilon$ can be 
approximated by $t/B^2+CB$ where $B$ is the GS extension and  
is given by $\partial \epsilon/\partial B=0$. This yields 
$B\propto (C/t)^{1/3}$ and $\epsilon \propto (U^2t/L^4)^{1/3}$. 
Since the $2d$-potential 
$v(\delta{\bf r})$ is separable, one eventually finds: 
\begin{equation}
E_{vib}^{(PSC)} (r_s,N=3) \propto r_s^{-\beta}
\label{Vibration-3-PSC} 
\end{equation}
in rydbergs where $\beta=4/3$. As one can see, the PSC repulsion 
gives a higher exponent $\beta$ when $N=3$, which is inconsistent with 
the usual expansion \cite{wigner} in powers of $r_s^{-1/2}$ 
first proposed by Wigner. 

Using Eq. \ref{Vibration-3-PSC}, one gets from Criterion 3 $r_l^*$  
given by Eq. \ref{threshold} again, but with $\alpha=3$ for the 
PSC repulsion instead of $\alpha=4$ for the PRC repulsion. 
The PSC repulsion is somewhat unphysical and leads to stronger 
lattice effects, but provides an interesting check of the validity 
of our theory: The changes of $\Delta E_{Coul}^{(PSC)}$ 
and $E_{vib}^{(PSC)}$ are such that the different criterions give 
thresholds $r_l^*$ which are consistent.  

The dimensionless energy ratio $F_N(L,U,t)$ for the PSC repulsion 
is shown in Fig. \ref{Fig8} for even values of $L$, where 
the GS is a triangular ``molecule'' shown in the inset when 
$t/U \rightarrow 0$.  Again the curves scale up to the onset 
$r_l^*(L)$ given by Criterion 3. But the PSC repulsion gives rise 
to a different onset $r_l^*(L)$ than the PRC repulsion for $N=3$, 
since at intermediate $r_l$ one has $F_{N=3} \propto r_l^{2/3}$ for 
the PSC repulsion, and not $\propto r_l^{1/2}$ as for the PRC repulsion. 

Does this difference remain for larger values of $N$? Indeed the 
contribution of pairs $ij$ having the coordinates of their spacings 
$d_{ij}$ close to $D/2$, and responsible for the $r_s^{2/3}$ behavior 
when $d_{ij}$ is defined by Eq. \ref{distance1}, becomes a surface 
effect $\propto N$ compared to the bulk contribution $\propto N^2$ 
of the remaining pairs, yielding $\Delta E_{Coul}^{PSC} \approx AN/L^2 
+ BN^2/L^3$, where $A$ and $B$ are constant. For a fixed $L$ and 
increasing $N$, $\Delta E_{Coul}^{PSC} \rightarrow B N^2/L^3$ and following 
Criterion 1, the conventional $r_s^{1/2}$ expansion for $F_{N}$ should be 
valid for the PSC repulsion too. Therefore, large periodic square lattices 
should exhibit a behavior independent of the choice of the long range part 
of the Coulomb repulsion when $N$ becomes large. Another possible choice 
is the Ewald repulsion obtained after summing over all the electrons 
present in the infinite repetition of the same finite square in the $x$ 
and $y$ directions. For a small number $N$ of electrons in a periodic 
square, these definitions are somewhat arbitrary. But to reach the 
thermodynamic limit, the PSC repulsion is less appropriate than the 
PRC or Ewald repulsions, since it gives larger finite $N$ effects.

 Nevertheless, the following relations for the lattice threshold, 
the continuum zero point energy of the crystalline oscillation and 
the characteristic scale of the Coulomb energy respectively:  
\begin{eqnarray}
r_l^* &\propto& L^{\alpha} \\
E_{vib} &\propto& r_l^{-\beta} \\
\Delta E_{Coul} &\propto& UL^{-\gamma} 
\end{eqnarray}
remain valid independently of the used definition of the Coulomb 
repulsion in the periodic square lattice, 
\begin{equation}
\alpha= \gamma+1 \ \ , \alpha =\frac{2}{2-\beta}
\end{equation} 
between the exponents.

\end{document}